# The Physical Nature of "Giant" Magnetocaloric and Electrocaloric Effects

**Vitaly J. Vodyanoy[1] and Yuri Mnyukh[2]**

[1]Biosensors Laboratory, Auburn University, Auburn Alabama USA
[2]Chemistry Department and Radiation and Solid State Laboratory, New York University, New York, NY USA,
*yuri@mnyukh.com*
(Dated: June 3, 2013)

**Abstract** Physical nature of "giant" magnetocaloric and electrocaloric effects, MCE and ECE, is explained in terms of the new fundamentals of phase transitions, ferromagnetism and ferroelectricity. It is the *latent heat* of structural (nucleation-and-growth) phase transitions from a normal crystal state to the orientation-disordered crystal (ODC) state where the constituent particles are engaged in thermal rotation. The ferromagnetism or ferroelectricity of the material provides the capability to *trigger* the structural phase transition by application, accordingly, of magnetic or electric field.

## 1. Introduction

*Magnetocaloric effect* (MCE) is the heat emanated or absorbed when magnetic field is applied to magnetic material. In principle, some thermal effect can be produced by any magnetic material: the applied magnetic field makes its structure unstable, creating conditions for changing the currently existing directions of its spins toward the direction of magnetic field. If the structural rearrangement occurs, the energy gain turns to heat. The effect, however, is small. The term MCE usually designates the much stronger effect, observed when a *phase transition* in the material is involved.

Oliveira and von Ranke (O&R) have published a comprehensive review on MCE, with 238 references [1], which is a good representative of the theoretical literature on the subject. Their conclusion that "underlying physics behind the magnetocaloric effect is not yet completely understood" was an understatement. In fact, the whole search for physical origin of the phenomenon was misdirected. As a result of the conventional incorrect interpretation of ferromagnetic state and solid-state phase transitions in general, the MCEs were erroneously ascribed to changing in magnetization. We will show that, instead, it is rooted in changing of the *crystal structure*.

The purpose of the present article is to reveal the physical nature of the "giant" MCE and its electrical counterpart *Electrocaloric Effect* (ECE). Since the previous research was misdirected, so were the efforts to find the most effective refrigerants for technologically sound magnetic refrigeration. In a seaming accordance with the term "magnetocaloric", the efforts were based on the belief that MCE is a *change in magnetic entropy* that can be estimated from magnetization measurements.

The MCE temperature is frequently reported as located in vicinity of magnetic phase transitions. The transitions were identified either as first order, or second order, or structural, or magnetic, or magnetostructural. There are descriptions of MCE as resulted from a "randomization of domains" at the Curie temperature. But the randomization process was not sufficiently understood either. Besides, the Curie temperature assumes a *second order* phase transition and, therefore, zero hysteresis, but hysteresis is known as a problem in the magnetic refrigeration technique. The MCEs in most experimental works were related to first-order phase transitions. The real molecular mechanism of the first order phase transitions could become a clue to the MCE origin, but they were treated only in a theoretically-formal manner, basically as "jumps" in the physical properties. The physical mechanism of these phase transitions remained in the dark.

The nature of MCE can be revealed only in terms of the real physics of a ferromagnetic state and solid-state phase transitions. The reader can find it in the book [2] and articles [3-8]. The next several sections are a background necessary for the explanation of the MCE and ECE that follows.

## 2. Classification of Phase Transitions by First- and Second-Order [2,3,9]

The classification of phase transitions by *first order* and *second order* has been taken for granted in the solid-state physics. It overwhelms the O&R review where it is used simply as statements "[X] is undergoing a first order phase transition", and "[Y] is undergoing a second order phase transition". In that sorting out, the "first order" were mentioned 76 times, and the "second order" 24 times. The only mentioned criterion was whether the entropy is a continuous or discontinuous function of temperature and magnetic field. If the former, it is a second-order transition, if the latter, it is a first order transition. It was not specified what *physical* process stays behind each of these two names. The theories suggested by O&R were not applicable to first order phase transitions. What makes phase transitions to be first or second order remained unknown.

The problem of the first/second order classification has been detrimental to solid-state physics for many decades. Presently it has simple solution: it should never be put forward. Soon after *second order* phase transitions were theoretically proposed in 1930th, M. von Laue and other prominent physicists rejected the possibility of their existence on thermodynamic reasons. In disregard of the objections, L. Landau introduced his theory of *second order* phase transitions, suggesting that they "may also exist". An analysis of the issue has led us to conclude that in reality *they do not exist*. The Landau's examples of second-order phase transitions turned out first order, as did the ferromagnetic transitions in Fe, Co and Ni, as do the transitions in all ferromagnetics and ferroelectrics, as do all "order-disorder" phase transitions. Not a single well-documented second-order phase transition exists. All current "second order phase transitions" are classified superficially and will ultimately be re-classified, so the classification itself will be *de facto* nullified. *All solid-state phase transitions materialize by a nucleation-and-growth rearrangement of the crystal structure.* This process is the most energy-efficient, requiring energy to relocate one molecule at a time, and not the myriads molecules at a time as a cooperative (second order) process requires.

## 3. Ferromagnetic State and Phase Transitions [5,10]

It has been experimentally established that MCE is tightly bound to ferromagnetic phase transition. All the current literature on the subject, including the O&R article, is based on the idea that MCE resulted from a "change in magnetic entropy". The ferromagnetic transition is deemed to be a change in the magnetic order in the same crystal lattice, even when the transition is identified as first order.

This interpretation of ferromagnetic phase transitions made discovery of the MCE origin impossible. Spin interaction in ferromagnetic material must be very strong in order to infer the "change in magnetic entropy" large enough to fit the largest MCEs actually observed. The Heisenberg's quantum-mechanical theory of ferromagnetism seemingly provided that strong interaction, called *electron exchange interaction*. The theory was to explain why a ferromagnet is stable, while magnetic interaction of its spins, according to calculations, is not. The theory has been taken for granted, even though its initial verifications had to prevent its acceptance. The verifications produced a *wrong sign* of the exchange forces. In other words, the stability of ferromagnetic state would *decrease* rather than increase. Besides, the theory failed in many other respects. The sign problem was later carefully reexamined and found unavoidable. It was predicted that the "neglect of the sign may hide important physics" [11].

The predicted important physics resides in the power of crystal field. Its ability of imposing one or another magnetic order was overlooked. There is no need in the additional spin interaction to explain ferromagnetism. The real magnetic interaction constitutes only a few percentage points of the total free energy of a ferromagnetic crystal, the main part of it being the energy of crystal chemical bonding. The ferromagnetic crystal is stable *in spite* of the destabilizing effect of the magnetic spin interaction [5,12].

For years ferromagnetic phase transitions were regarded "structureless" by theoretical physicists. However, in time, experimental evidence was mounting that change in magnetic ordering in many cases is "accompanied by" a change in crystal structure. Such phase transitions are presently called "magnetostructural". The history of the matter needs in following two corrections: 1. Not only *many*, but *all* ferromagnetic phase transitions are "magnetostructural", and 2. The 'magnetic' and 'structural' roles in ferromagnetic phase transitions are actually different, considering that it is not the new magnetic ordering that brings about change of the structural change, as previously believed, but quite the reverse.

In relation to the search for the cause of MCE, it can be concluded that the thermal effect during a ferromagnetic phase transition comes from the *latent heat of the structural rearrangement*, while the contribution from the magnetic reordering is minor. Yet, that latter contribution might still be sufficient to determine which of the two competing crystal structures is preferable under given conditions.

## 4. Hysteresis, Nucleation and Range of Transition [6,13]

Hysteresis is always a hurdle in the way of practical realization of magnetic refrigeration. This fact is an indication that physical origins of MCE and hysteresis are closely related. Therefore, the statements by O&H "The present discussion does not consider the hysteresis effect" and "A microscopical description of the hysteresis effect on the magnetocaloric properties is still lacking" had to question their theory describing hysteresis-free second-order phase transitions. But hysteresis is always found. Its observation, even a smallest one, *binds the MCE to the process of nucleation and growth*.

At present, hysteresis has a detailed physical explanation. Not entering into all details, it is as follows. Magnetic ordering results from a *structural* rearrangement during ferromagnetic phase transition. The rearrangement materializes by *nucleation and growth*. The *nucleation* is not the classical fluctuation-based process described in textbooks. In a given crystal it is pre-determined. The nuclei are located in specific crystal defects - microcavities of a certain optimum size. These defects already contain individual information on the temperatures $T_n$ of their activation. The nucleation lags $\Delta T_n = T_n - T_o$ (at $T_o$ the free energies of the polymorphs are equal) are inevitable and not the same in different defects. These lags are the only cause of the *structural hysteresis*. Since each crystal structure determines its pattern of magnetic ordering, this structural hysteresis is a magnetic hysteresis as well. It is the hysteresis in phase transitions. The hysteresis of (re)magnetization by applied magnetic field has the same cause, namely, nucleation lags of the underlying structural rearrangements [10].

Considering that almost all real systems feature multiple nucleation, the phase transitions spread over a *range of transition* − temperature or magnetic depending on the driving parameter. The range as a whole is a subject to hysteresis, which means that the ranges in the direct and the reverse runs will not overlap.

## 5. Heat Capacity and λ-Anomaly [7,14]

MCE is victim of a mistake permeated all the literature where a heat capacity over the temperature region of phase transition is reported. Assigning the peaks like one in Fig. 1 to a heat capacity is erroneous. As an example, more than 30 such "specific heat" peaks were reproduced in [15]. These peaks have long history. After first recorded in 1922, they were named "heat capacity λ-anomalies" and regarded to be a feature of second order phase transitions. Their shape was meticulously analyzed in the efforts to figure out the nature of "critical phenomena". The "λ-anomalies" remained a mystery even after discovery that they "also" appear in first-order phase transitions. At that point, it seems, it had to become clear that the peaks are not an anomaly, but rather a *latent heat* of the structural phase transitions. That did not happen, possibly on the following reasons: (a) The first-order phase transitions were treated only formally, something like second-order phase transitions interrupted by "jumps", (b) The λ-peaks in the $C_P$ (T) plots in ferromagnetic phase transitions were ascribed to the magnetic properties and treated as a "specific heat near the Curie transition", and (c) The adiabatic calorimetry typically used in the measurements has a usually unnoticed limitation. An adiabatic calorimeter permits measurements only upon heating. While it correctly shows the λ-peak being endothermic, it is incapable to reveal that the same peak will be exothermic (looking downward) and located at lower temperature in the cooling run. This can be revealed with a differential scanning calorimeter and thus prove that the mysterious "anomaly" is simply a latent heat of the phase transition.

## 6. Crystal State with Rotating Molecules [16]

It is presently known that large MCEs are associated with the "order-disorder" phase transitions between magnetically ordered ferromagnetic and magnetically disordered paramagnetic states. As to the latter, the curiosity in the MCE research did not go farther of noting that spin orientations are randomized by thermal motion. While describing the MCE as change in magnetic entropy, it would be wise for the MCE research to recall about existence of orientation-disordered crystals with *thermal rotation of their molecules* and take a close look at that state.

This specific solid state was attracting a greater attention in the middle of the last century, culminating in the publication of the book [16], due to its novelty and hope to replenish the dwindling quantity of second-order phase transitions,. The substances revealing this state usually have rounded molecules. The state is characterized by the long-distance translation order in the molecular positions but not orientations owing to molecular thermal rotation. A rotational phase is always the highest on the temperature scale, right before melting. The rotation is never quite free, even when the long-distance orientational order is completely lost. With increasing temperature the molecular orientational disordering frequently occurs in several stages - from increased librations, to frequent jumps between certain discrete orientations, to 2-D rotation, to "free" rotation. However, even in the latter case the rotation is still hindered, retaining some degree of a short-distance orientational order.

All said about the whole molecules can be repeated when only their parts are subjected to thermal rotation. For example, in $NH_4Dy(SO_4) \cdot 4H_2O$ the following major stages were found [17]: (1) the ammonium ions are undergoing well defined librations at $95 < T < 200$ K, (2) their molecular motions at $200 < T < 275$ K are complex, probably a superposition of large-amplitude reorientational motion and small-amplitude librational fluctuations, and (3) they attain almost free rotation at $T > 275$ K.

All the order-disorder phase transitions in question, including those between the orientation-disordered stages, are first order, meaning they occur by nucleation and growth. They are among the solid-state phase transitions with the largest latent heat.

## 7. Paramagnetic State

Importance of correct interpretation of paramagnetic state goes far beyond of the MCE research. As for the literature on MCE, it is mentioned only as a state with orientation-disordered *spins* or, at best, as a state with the spins randomized by thermal motion. What is missing in that interpretation is that spins are not independent entities: they are bound to their atomic and molecular carriers (see Section 3). Therefore, a common statement like "thermal energy overcomes the interaction energy between the spins" is incorrect at the point that happens to be crucial for finding the MCE origin. The interaction energy between the spins is weak as compared to the molecular chemical bonding. It is the latter that the thermal energy must overcome to let the atomic-molecular spin carriers become rotating.

The truth of the matter is that paramagnetic crystals are the *orientation-disordered crystals (ODCs) with thermal rotation of their atoms and molecules*. The crystal restructuring in a *ferromagnetic − paramagnetic* phase transition should not be overlooked. As all *order − disorder* phase transitions, they occur by nucleation and growth and, therefore. involve latent heat.

## 8. Physical Nature of the "Giant" MCE in $Gd_5(Si_2Ge_2)$

The discovery of a "giant" MCE in $Gd_5(Si_2Ge_2)$ by Pecharsky and Gschneidner (P&G) [18] was a milestone toward development of the new refrigeration technology. But the discovery, best represented by the plot in Fig. 1, left the MCE without correct explanation of its physical nature. The MCE, represented by the λ-peak at the temperature of first-order phase transition 276 K, was ascribed to the " magnetic entropy change".

In terms of properties of the real solid-state phase transitions described in Sections 2 - 7, the nature of the "giant" MCE in question becomes almost self-explanatory. There is no basis for that MCE to be a "change of magnetic entropy". It is of *crystal-structural* origin, being the *latent heat of the structural phase transition* (by nucleation and growth) in the magnetic material. The MCE appears on the experimental heat capacity curves $C_P(T)$ as a peak that used to be a "λ-anomaly" indicative of the second-order phase transition. In the MCE studies these peaks were turned to a "heat capacity due to change in magnetic entropy". In fact, they are equally observed in non-magnetic materials and are proven [14] to be the *latent heat* of nucleation-and-growth phase transitions. In case of any doubts this can be verified by a differential scanning calorimetry: the λ-peak in the cooling run will emerge exothermic (looking downward) and, at that, at a lower temperature due to hysteresis.

It becomes now possible to resolve another contradiction in the interpretation by P&G [18] of their giant MCE (see Fig.1). To be in accord with the empirically established rule, this giant MCE had to be ascribed to the small "anomaly" (arbitrarily claimed to mark a second order phase transition) produced by the *ferromagnetic − paramagnetic* (FM−PM) transition at 299 K − the type of phase transition where spin randomization or ordering occurs and the "magnetic entropy change" should be maximal. But the MCE was instead presented by the λ-peak produced by the (FM−FM) transition at 276 K, the transition type where both phases were found magnetically ordered.

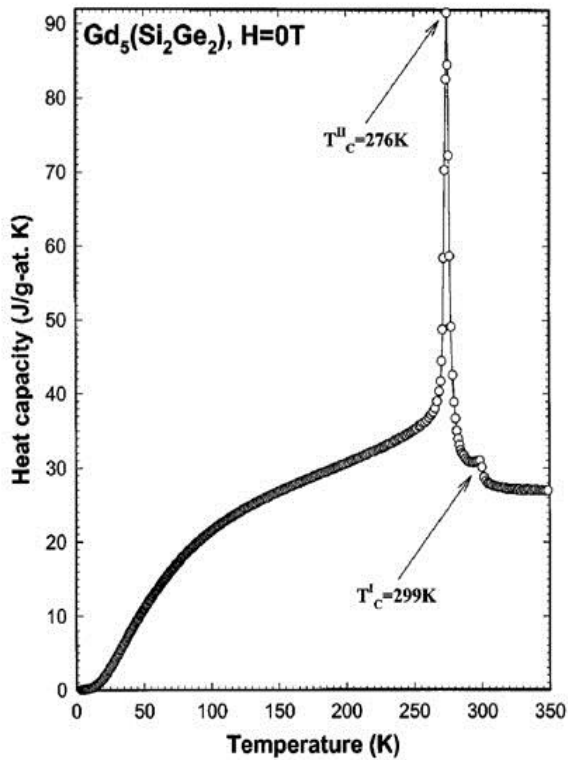

FIG. 1. The zero magnetic field heat capacity of $Gd_5(Si_2Ge_2)$ from 3.5 to 350 K. The arrows point to heat capacity anomaly due to a second order paramagnetic ↔ ferromagnetic (I) transformation at 299 K and a first order ferromagnetic (I) ↔ferromagnetic (II) transition at 276 K. (This is the original figure capitation reproduced from [6]. Used with permission from APS Associate Publisher).

To explain this, we will trace the structural changes from the higher temperatures down. Molecular rotation in the ODCs is known to be hindered rather than quite free (see Section 6). The PM→FM phase transition at 299 K is $ODC_1 \rightarrow ODC_2$. The increased hindrance to molecular rotation with decreasing temperature turns the isotropic distribution of molecular orientations (and their spins) to anisotropic, thus converting the paramagnetic $ODC_1$ to a ferromagnetic $ODC_2$. Since it is still a *rotational − rotational* transition, only small change in the total crystal energy is involved (small latent heat peak at 299 K). The phase transition at 276 K features a giant MCE, for it is a transition of the rotational $ODC_2$ (FM) phase to the normal crystal with fixed molecular orientations. The latent heat of this phase transition is represented by the area of the λ-peak superimposed on the true $C_P(T)$ curve.

The magnetic change does not contribute essentially into the MCE. The magnetic component of the crystal free energy is small, being only sufficient to change the energy balance toward or against the ODC phase and *trigger* the phase transition when magnetic field is applied. The phase transition can also be triggered by change of temperature T, and the same latent heat could be named "TCE", or it can be triggered by pressure P and named "PCE". In fact, the λ-peak in Fig. 1 *is* the TCE. But using a temperature change as a trigger in the refrigeration technique is impractical. However, application of electric field to *ferroelectric - paraelectric* phase transitions is not impractical. It can produce the *electrocaloric effect* (ECE) quite analogous to the MCE.

## 9. Electrocaloric Effect (ECE)

As all solid-state phase transitions, *ferroelectric - paraelectric* phase transitions materialize by nucleation and growth. If triggered by application of electric field, their latent heat becomes *an electrocaloric effect ECE*. The ECE has long scientific history [19]. Accounting for its origin had turned out even more problematic, considering that nothing analogous to the "magnetic entropy change" and "electron exchange field" is applied to dielectrics. No sound explanation of the ECE exists. Yet, the ECEs comparable to the "giant" MCEs were attributed to the large *polarization change*. "near or above" the *ferroelectric → paraelectric* transition. Though random orientation of the electric dipoles in the paraelectric phase was recognized, the understanding was missing that this phase is an ODC where (dipolar) *molecules* are engaged in thermal rotation. It is the energy of the crystal restructuring, involving a conversion to thermal rotation of the constituent molecules, that makes accounting for the "giant" ECEs easy and identical to that for MCE.

## 10. Inverse MCE

Historically, the name "magnetocaloric effect" MCE was given to the *heating* upon application of magnetic field H. We now explain this as exothermic effect of the *disordered* (DIS) → *ordered* (ORD) structural phase transition caused by the orienting action the applied magnetic field H exerts on the disordered spins.

This left the name "Inverse MCE" to the endothermic effect also observed sometimes upon H application. Obviously, it results from the ORD → DIS transition. In other words, application of H under some circumstances would destabilize the ORD phase more than it does to the DIS phase. In one such situation the action of the applied H can sufficiently strengthen spin interaction in the ferromagnet where that interaction, by itself is destabilizing (see Section 7). Another possible case is the H destroying antiferromagnetic order in the (AFM)ORD → DIS. transition.

## 11. Conclusions

Hopefully, the physical nature of the magnetocaloric and electrocaloric effects is not an enigma any more. It is always good to understand the origin of a physical phenomenon, but this is especially true in attempts to use it in practice. In case of MCE and ECE it may be now expedient to modify the direction of the search for best refrigerants. In the erroneous belief that magnetic properties are responsible for the *size* of MCE, a disproportional attention was paid to the magnetic measurements. In fact, ferromagnetism provides only the ability to *trigger* phase transitions by magnetic field. It may be prudent to shift a part of that attention toward analyzing the structure and properties of the orientation-disorder crystals (ODC) with thermal molecular rotation.

As for the ECE, it should be useful to know that search for its separate explanation can be canceled. The physical origin of the both MCE and ECE is the same, and it is *latent heat* of the structural phase transitions.

The function of the applied electric field is quite analogous to the applied magnetic field in the MCE case, namely, to *trigger* those phase transitions.